\documentclass[12pt]{article}
\pdfoutput=1
\usepackage{epsfig}
\usepackage{amsfonts}
\usepackage{amssymb}
\usepackage{xcolor}
\usepackage{graphicx} % Required for inserting images
\usepackage[cp1251]{inputenc} %кодировка
\usepackage[T1]{fontenc}   % Латинские символы
\usepackage[english]{babel} %для нормального шрифта
\usepackage{indentfirst} %Красная строка первого абзаца
\usepackage{amsmath} %чтобы работала нумерация
\usepackage{float} %чтобы работала нумерация
\usepackage{caption}  %заголовки плавающих объектов

\begin{document}

\begin{center}

{\bf \Large  Asymptotic Freedom of V-A Fermi Interaction} \vspace{1.0cm}

{\bf \large A. T. Borlakov and D. I. Kazakov} \vspace{0.5cm}

{\it Bogoliubov Laboratory of Theoretical Physics, Joint
Institute for Nuclear Research, Dubna, Russia}
\vspace{0.5cm}

\abstract{We consider the V-A Fermi interaction and apply an earlier developed method  for summing up the leading  asymptotics for scattering amplitudes in non-renormalizable theories. We  consider the amplitude of fermion-antifermion scattering and derive the corresponding RG equation that sums the leading logarithmic contributions just like in renormalizable models. Numerical solution of this equation in the asymptotic regime $s\sim t\sim u \sim E^2 \to \infty$ leads to amplitude logarithmically decreasing with energy, thus restoring the unitarity violated at the tree level.}
\end{center}

\section{Introduction}

Consider the Fermi theory described by the Lagrangian
\begin{equation}
{\cal L}=i \bar \Psi \hat \partial \Psi - \frac{G}{\sqrt{2}} (\bar \Psi\hat{\cal O}_i \Psi) (\bar \Psi\hat{\cal O}_i \Psi),
\end{equation}
where the operator $\hat{\cal O}=\gamma^\mu(1-\gamma^5)/2$ and $G$ is the Fermi constant.  It is the low-energy effective theory of weak interactions replaced in the Standard Model by the gauge theory with intermediate vector bosons.
 The reason for this is the  violation of unitarity already above the EW scale, which arises due to an increase in the amplitude of the 4-fermion interaction with energy at the tree level. It is also wellknown that Fermi theory is non-renormalizable starting with the first loop.  This means that the improvement of high-energy behaviour with the standard RG technique is not  applicable here.

However, in a series of papers~\cite{BKTV,KBTV,KBBTV} we showed how to work with non-renormalizable theories. We constructed the corresponding renormalization group equations that allow us to sum up  the leading logarithms contributing to the scattering amplitude in the same way as in renormalizable theories. It was demonstrated that radiative corrections summed over all orders of PT can significantly change the behaviour of the amplitude and can lead to a decrease in the amplitude with energy.

The advocated approach is based on the BPHZ $R$-operation~\cite{BSh,Hepp,Zimmer} and Bogoliubov-Parasiuk theorem~\cite{BP,APZ}. The key point is that after applying the incomplete $R$-operation, the so-called $R'$-operation, the counter terms appear to be always local, i.e. contain a finite number of derivatives in coordinate space or at most polynomials in momentum space.
The requirement of locality imposes severe restrictions on the structure of the counter terms and allows one to
write down the recurrence relations connecting the counter terms in subsequent orders. These recurrence relations can be promoted to the differential equations  for the generating functions which are nothing else but the RG equations. Solutions to these equations give the sum of the leading, subleading, etc. divergences in all orders of PT.

Here we apply the developed formalism to the V-A Fermi theory. We consider the  amplitude of fermion - antifermion scattering  and derive the RG equation to sum up the leading logarithmic asymptotics.  In doing so, we are aware of the fact that in non-renormalizable theories  one  has infinite arbitrariness in the subtraction procedure that makes the theory hardly predictable. However, the leading order terms are independent of this arbitrariness. So it does not matter how one subtracts the divergences, one can do it, for instance, in a minimal way. Thus, without  really defining the full theory, which is of course an ambitious task, we can still get an expression  for the scattering amplitude in the leading order.

\section{Fermi theory in spinor helicity formalism}

\begin{figure}[htb]
\begin{center}
\includegraphics[scale=0.6]{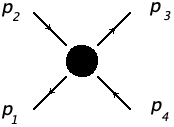}
\caption{The Feynman diagram for the fermion-antifermion scattering amplitude.  All momenta are incoming.
}
\label{amp}
\end{center}
\end{figure}

Our aim is to calculate the amplitude of fermion-antifermion scattering shown in Fig.\ref{amp}.

For the analysis of the four-fermion theory it is very convenient to use the spinor-helicity formalism~\cite{Elvang, 1001, Schwinn}.
Then, in the V-A case, the tree amplitude has only one spinor structure

\begin{equation}
A_0= \frac 14 \langle 13\rangle [42], \label{treeA}
\end{equation}

where the factor $\frac 14$  takes into account the equivalence of external states.
All radiative corrections are proportional to the same structure, so we calculate the ratio to the tree amplitude and omit the common factor.

In each order of PT, the amplitude in its turn is given by the sum of s-,t- and u-channel contributions
\begin{equation}
   A_n(s,t,u) = S_n(s,t,u)+T_n(s,t,u)+U_n(s,t,u).
    \label{Ak}
\end{equation}
We use dimensional regularization to handle UV divergences and calculate integrals in $4-2\epsilon$ dimensions. Then the calculation of one-loop diagrams in the s-,t- and u-channel gives, respectively (singular parts only),
\begin{equation}
 S_1(s,t,u) =-\frac{8s}{3\epsilon},\ T_1(s,t,u) =-\frac{8t}{3\epsilon},\ U_1(s,t,u) =\frac{8u}{\epsilon} .
 \label{1stu}
\end{equation}
In higher orders, we calculate the leading divergences (leading poles in $\epsilon$). The coefficients of the leading poles are equal to those of the leading logarithms, which is the aim of  our calculations. We first do it by explicitly calculating the diagrams using the spin-helicity formalism, which essentially simplifies the task. The resulting leading poles up to three loops are~\cite{Borlakov2}:

\begin{equation}
\begin{aligned}
 A_1&=-\frac{8}{3\epsilon} (s + t - 3 u) ,\\
A_2&=\frac{64}{9\epsilon^2} (s^2 + t^2 + 6 u^2) ,  \\
A_3&=-\frac{256}{405\epsilon^3}(37 s^3 + 37 t^3 - 6 s^2 u - 6 t^2 u - 318 u^3). 
\label{three}   
\end{aligned}
\end{equation}

\begin{figure}[htb]
\includegraphics[scale=0.45]{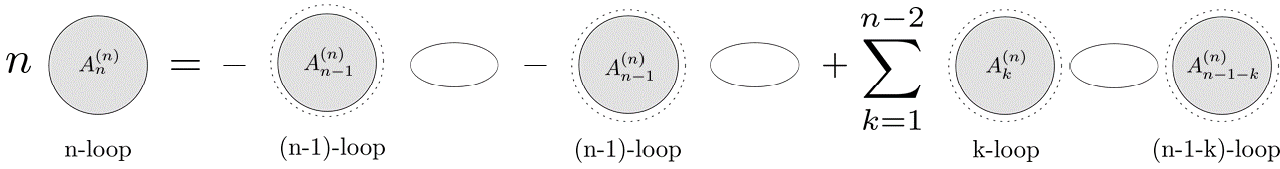}
\caption{The schematic form of the recurrence relation for the four-point function. The bubbles surrounded by the dotted lines denote the corresponding counter terms}
\label{rec}
\end{figure}

To calculate higher loop contributions, we write down the recurrence relations  first  derived in \cite{Borlakov}.
They connect the nth order coefficients  with the lower order ones and have to be written separately for each channel.  Schematically, they are shown in Fig.\ref{rec}

\noindent and have the following analytic form  (we omit the obvious factor $1/\epsilon^n $):
\begin{equation}
\begin{aligned}
          nS_n(s,t,u) = &-8s\!\int_{0}^{1}\!\!\! dx \sum_{k=0}^{n-1}\sum_{p=0}^{k}\frac{[s(-s-u)]^p[x(1-x)]^{p+1}}{p!(p+1)!(p+2)^{-1}}\times\\
        &\times\frac{d^p A_{k}(s,-s-u',u')}{du'^p}\frac{d^p A_{n-1-k}(s,-s-u',u')}{du'^p}\mid_{u'\rightarrow-sx},
\end{aligned}
    \label{Sn}
\end{equation}
\begin{equation}
\begin{aligned}
          nT_n(s,t,u) = &-8t\!\int_{0}^{1}\!\!\! dx \sum_{k=0}^{n-1}\sum_{p=0}^{k}\frac{[t(-t-u)]^p[x(1-x)]^{p+1}}{p!(p+1)!(p+2)^{-1}}\times\\
        &\times\frac{d^p A_{k}(-t-u',t,u')}{du'^p}\frac{d^p A_{n-1-k}(-t-u',t,u')}{du'^p}\mid_{u'\rightarrow-tx}  ,
\end{aligned}
    \label{Tn}
\end{equation}
\begin{equation}
\begin{aligned}
         nU_n(s,t,u) &= 8u\!\int_{0}^{1}\!\!\! dx \sum_{k=0}^{n-1}\sum_{p=0}^{k}\frac{[u(-s-u)]^p[x(1-x)]^{p+1}}{p!(p+1)!(p+3)^{-1}}\times\\
         &\times\frac{d^p A_{k}(s',-s'-u,u)}{ds'^p}\frac{d^p A_{n-1-k}(s',-s'-u,u)}{ds'^p}\mid_{s'\rightarrow-ux}\\ 
         &+8u\!\int_{0}^{1}\!\!\! dx \sum_{k=0}^{n-1}\sum_{p=0}^{k}\frac{[u(-t-u)]^p[x(1-x)]^{p+1}}{p!(p+1)!(p+3)^{-1}}\times\\
         &\times\frac{d^p A_{k}(-t'-u,t',u)}{dt'^p}\frac{d^p A_{n-1-k}(-t'-u,t',u)}{dt'^p}\mid_{t'\rightarrow-ux}.
\end{aligned}
    \label{Un}
\end{equation}

Several comments are in order: the recurrence relations are written for the counter terms. In the leading order, they coincide with the leading divergences of the diagrams up to a factor $(-)^n$ where $n$ is the order of PT. The divergent parts are polynomials of momenta
as one can see, for instance,  from eq.(\ref{1stu}). These momenta are external for the counter-terms but might be internal for the whole diagram, so one has to integrate over them in the remaining one-loop diagram shown in Fig.\ref{rec}. This explains the appearance of the integral over the Feynman parameter $x$ in recurrence relations (\ref{Sn}-\ref{Un}).

These recurrence relations  allow one to calculate the leading divergences in any order starting with the one-loop  contribution (\ref{1stu}).
One can check that using the  one-loop expressions (\ref{1stu}) and substituting them into recurrence relations (\ref{Sn}-\ref{Un}), one can reproduce the higher order terms (\ref{three}).

The recurrence relations can be converted into the differential equation for the generating function
\begin{equation}
A(s,t,u)=\sum_{n=0}^\infty A_n(s,t,u)(-z)^n, \ \ \ z=\frac{\bar G}{\epsilon},
\end{equation}
where $\bar G= G/\sqrt{2}/16\pi^2$ is the standard loop expansion parameter.
It takes the form of the integro-differential equation for the scattering amplitude
\begin{equation}
    \begin{aligned}
            &-\frac{dA(s,t,u)}{dz}=
            \\
            &=\!-8s\!\int_{0}^{1}\!\!\! dx \sum_{p=0}^{\infty}\frac{[s(-s-u)]^p[x(1-x)]^{p+1}}{p!(p+1)!(p+2)^{-1}}\left(\frac{d^p A(s,-s-u',u')}{du'^p}\right)^2
            \\
            &\hspace{1em}-8t\!\int_{0}^{1}\!\!\! dx \sum_{p=0}^{\infty}\frac{[t(-t-u)]^p[x(1-x)]^{p+1}}{p!(p+1)!(p+2)^{-1}}\left(\frac{d^p A(-t-u',t,u')}{du'^p}\right)^2
            \\
            &\hspace{1em}+\!8u\!\int_{0}^{1}\!\!\! dx \sum_{p=0}^{\infty}\frac{[u(-s-u)]^p[x(1-x)]^{p+1}}{p!(p+1)!(p+3)^{-1}}\left(\frac{d^p A(s',-s'-u,u)}{ds'^p}\right)^2
            \\
            &\hspace{1em}+\!8u\!\int_{0}^{1}\!\!\! dx \sum_{p=0}^{\infty}\frac{[u(-t-u)]^p[x(1-x)]^{p+1}}{p!(p+1)!(p+3)^{-1}}\left(\frac{d^p A(-t'-u,t',u)}{dt'^p}\right)^2.
            \\
    \end{aligned}
    \label{EqA}
\end{equation}
Solution to this equation sums up the leading contributions to all orders of PT.
Since an analytical solution of such a complex integro-differential equation is not possible, we turn to numerical methods. 

\section{Numerical solution of RG equation in asymptotic regime}

To perform a numerical solution, we consider the amplitude in the high-energy asymptotic regime when $s = 4E^2, t=u=-s/2$  in the center-of-mass frame and $E\to \infty$.  Since the coefficients of the leading logarithms $\log^n\mu^2$ coincide
with those of the leading poles $1/\epsilon^n$,  to get the perturbative expansion of the amplitude, one has to replace everywhere $\bar G/\epsilon \to -\bar G \log(s/\mu^2)$.   Moving on to the dimensionless variable $y=\bar G s\log (s/\mu^2)$, we notice that the series in this case is sign alternating and have the form
\begin{equation}
A(s,t,u)= 1 - \frac{16}{3}y+\frac{176}{9}y^2-\frac{19424}{405}y^3+ ...
    \label{ampPT}
\end{equation}
Thus, based on the analysis of a finite number of terms of the series, it is impossible to make a conclusion about the asymptotic behaviour of the amplitude. Therefore, the numerical solution of the generalised renormalization group equation, which sums an infinite number of terms of perturbation theory, becomes important.
%
%In order to find the dependence of the scattering amplitude on energy,
%we use the fact that the

In order to make the obtained equations more suitable for numerical analysis, it is  convenient to replace the operation of infinite summation  in (\ref{EqA}) with integration.  To do this, we use the following trick: starting with the Taylor expansion of some function
\begin{equation}
f(A+Be^{\pm i\tau})=\sum_{k=0}^\infty \frac{B^k e^{\pm i\tau k}}{k!}\frac{d^kf(A)}{dA^k}\quad.
\end{equation}
and keeping in mind the orthogonality condition for periodic functions
\begin{equation}
   \frac{1}{2\pi} \int_{-\pi}^\pi e^{i\tau n}e^{-i\tau m}d\tau=\delta_{mn}\quad,
\end{equation}
as well as the useful formula
\begin{equation}
    \int_0^1 d\xi (1-\xi)^k\frac{\xi^p}{p!} = \frac{k!}{(p+k+1)!}\quad,
\end{equation}
one can obtain the following relation:
\begin{equation}
    \sum_{p=0}^{\infty} \frac{(BC)^pk!}{p!(p+k+1)!}\left[\frac{d^p f(A)}{dA^p}\right]^2=\frac{1}{2\pi}\int_{-\pi}^\pi \!\!\!d\tau \!\int_0^1\!\!\!d\xi(1\!-\!\xi)^kf(A\!+\!e^{i\tau}B\xi)f(A\!+\!e^{-i\tau}C)\quad.
    \label{BC}
\end{equation}
This relation allows one to get rid of an infinite sum in eqs.(\ref{EqA}) and evaluate the remaining integrals numerically.

\begin{figure}[!htb]
\center{\includegraphics[scale=0.7]{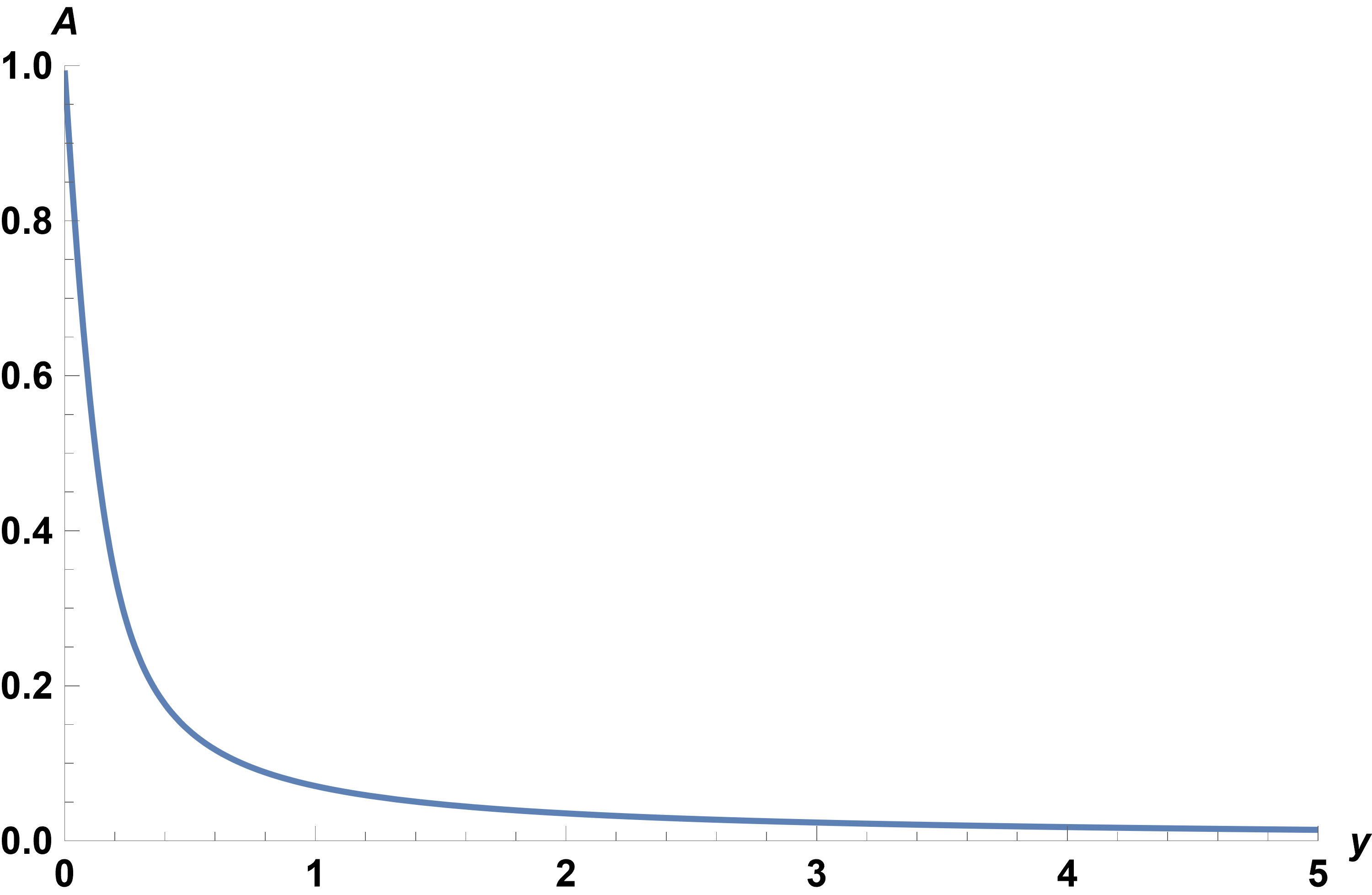}}
\caption{The function $A(y)$ as a numerical solution of equation (\ref{EqA})}
\label{GAlla}
\end{figure}

The results of numerical solution of equation(\ref{EqA}) in terms of the dimensionless variable $y$ is  shown in Fig.~\ref{GAlla}.

As can be seen, the resulting solution demonstrates a rapid decrease in the amplitude at high energies.
We have obtained the following numerical approximations of the function $A(y) = 0.067y^{-0.987}$  at the $R^2=0.9973$ coefficient of determination.

This leads to the following asymptotic behaviour of the function $A(y)$  at  high energy scale:
\begin{equation}
  A(y) \approx \frac{k}{y}=\frac{k}{\bar G s\log (s/\mu^2)}, \ \ k\approx1/15.
  \label{approx}
\end{equation}

\section{Discussion}

Thus, we see that summation  of the leading logarithmic corrections to the four-fermion scattering amplitude in V-A Fermi theory  significantly changes  the behaviour of the amplitude.  While the terms of the PT series  increase with energy, the solution of the renormalization group equation is characterized by asymptotically free behaviour!

The latter circumstance has a far-reaching physical consequence for the V-A Fermi theory. Indeed,
the differential  scattering cross-section in the tree approximation is equal to:
\begin{equation}
     \frac{d\sigma}{d\Omega}=\frac{(G/\sqrt{2})^2u^2/16}{16\pi^2s}=\frac{(G/\sqrt{2})^2s/64}{16\pi^2}\quad,
     \label{xsec}
\end{equation}
where we have taken into account the value of the tree level amplitude (\ref{treeA}) $|A_0|^2=\frac{1}{16}u^2$. This expression  violates the unitarity condition \cite{peskin, Schwartz}
\begin{equation}
     \frac{d\sigma}{d\Omega}\leq \frac{1}{4\pi^2s}\label{bound}
\end{equation}
at  $s>16\sqrt{2}/G\sim (1500\  GeV)^2$ for $G=10^{-5}/m_p^2$.\footnote{Note that this bound  depends on the kinematics and is slightly different from that for a forward scattering amplitude, which can be limited by a partial wave unitarity bound~\cite{Schwartz}.}   At the same time, summation of  the leading  logarithms leads to asymptotic behaviour of the amplitude that is described with fairly good accuracy by expression
\begin{equation}
 \frac{d\sigma}{d\Omega}\approx\frac{(Gs/8\sqrt{2})^2}{16\pi^2s}\frac{k^2}{(G/\sqrt{2}\cdot16\pi^2)^2s^2\,\log^2(s/\mu^2)}=\frac{C^2}{s\log^2(s/\mu^2)}<\frac{1}{4\pi^2s}\quad,
    \label{csF}
\end{equation}
where $C=\pi k/2 \sim 0.1$ is a small constant. Therefore, the unitarity bound is now satisfied.
Thus, summation of the leading logarithms restores the unitarity in Fermi theory at high energy.

It is instructive to compare the expression obtained for the cross section~(\ref{csF}) with that obtained in the theory with an intermediate gauge boson. In the latter case, in the tree approximation, one has~\cite{Schwartz}:
\begin{equation}
    \frac{d\sigma}{d\Omega}=\frac{1}{16\pi^2s}\frac{g^4s^2}{[s-M^2]^2}\stackrel{s\gg M}{\longrightarrow}\frac{g^4}{16\pi^2s}\quad,
    \label{gws}
\end{equation}
In this case, the higher-order PT corrections are taken into account using the standard renormalization group method and result in replacing the coupling constant $g^2$ with an effective "running" coupling constant~\cite{GG}
\begin{equation}
g_{eff}^2(s)=\frac{g^2}{1+(\beta_0/16 \pi^2)g^2\log(s/M_Z^2)}\quad,
\end{equation}
where the constant $\beta_0$ depends on a number of weak-interacting particles and equals $\beta_0=19/6$ in the Standard Model.

\begin{figure}[ht]
\center{\includegraphics[scale=0.7]{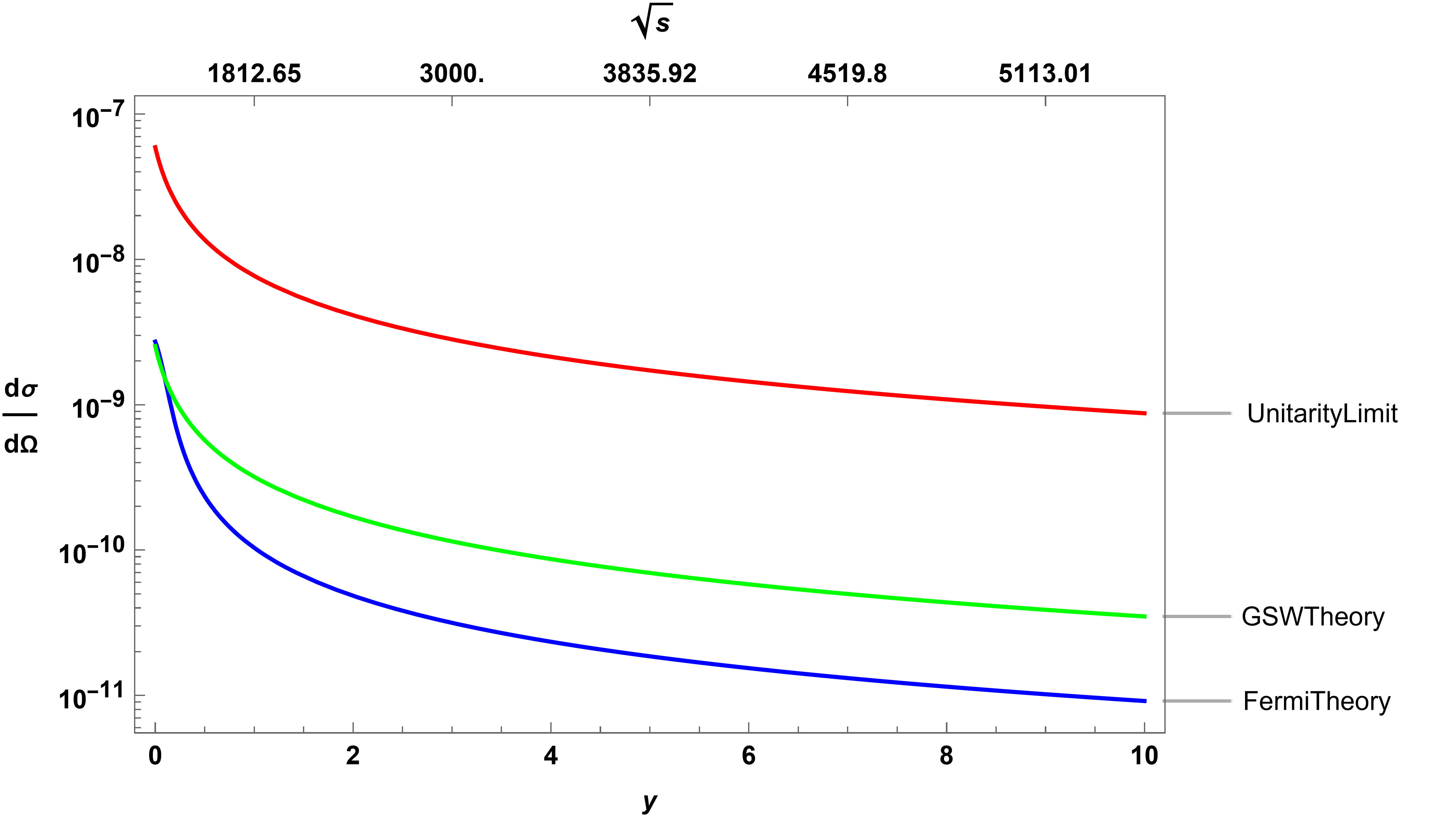}}
\caption{The differential cross-section in the SM and in Fermi theory as a function of $s$. The unitarity bound is also shown for reference.}
\label{Fit}
\end{figure}

For comparison, we plot the differential cross-section in Fermi theory with account for summation of the leading logarithms and in the Standard Model in Fig.\ref{Fit}. The unitarity bound is also shown for reference.

 One can see that the differential cross-section as a function of energy exhibits smooth behaviour
 and does not violate the unitarity bound everywhere. This demonstrates that non-renormalizable theories can be treated within perturbation theory improved by the renormalization group and can be applied to describe physical processes. Of course, the UV complete theory is still lacking but the obtained results are hopeful.

\section*{Acknowlegments}
The authors  are  grateful to the colleagues from BLTP for useful discussions.

\bibliographystyle{hunsrt.bst}
\bibliography{bibliography}

@article{BKTV,
    author = "Borlakov, A. T. and Kazakov, D. I. and Tolkachev, D. M. and Vlasenko, D. E.",
    title = "{Summation of all-loop UV Divergences in Maximally Supersymmetric Gauge Theories}",
    eprint = "1610.05549",
    archivePrefix = "arXiv",
    primaryClass = "hep-th",
    doi = "10.1007/JHEP12(2016)154",
    journal = "JHEP",
    volume = "12",
    pages = "154",
    year = "2016"
}

@article{KBTV,
    author = "Kazakov, D. I. and Borlakov, A. T. and Tolkachev, D. M. and Vlasenko, D. E.",
    title = "{Structure of UV divergences in maximally supersymmetric gauge theories}",
    eprint = "1712.04348",
    archivePrefix = "arXiv",
    primaryClass = "hep-th",
    doi = "10.1103/PhysRevD.97.125008",
    journal = "Phys. Rev. D",
    volume = "97",
    number = "12",
    pages = "125008",
    year = "2018"
}

@article{KBBTV,
    author = "Kazakov, Dmitry and Bork, Leonid and Borlakov, Arthur and Tolkachev, Denis and Vlasenko, Dmitry",
    title = "{High Energy Behavior in Maximally Supersymmetric Gauge Theories in Various Dimensions}",
    doi = "10.3390/sym11010104",
    journal = "Symmetry",
    volume = "11",
    number = "1",
    pages = "104",
    year = "2019"
}

@book{BSh,
    author = "Bogolyubov, N. N. and Shirkov, D. V.",
    title = "{Introduction to the theory of quantized fields}",
    publisher = "Nauka, Moscow, 1957. English transl.: 
                 Introduction to the Theory of Quantized 
                 Fields, 3rd ed., New York, Wiley",
    year = "1980"
}

@article{Hepp,
    author = "Hepp, Klaus",
    title = "{Proof of the Bogolyubov-Parasiuk theorem on renormalization}",
    doi = "10.1007/BF01773358",
    journal = "Commun. Math. Phys.",
    volume = "2",
    pages = "301--326",
    year = "1966"
}

@article{Zimmer,
    author = "Zimmermann, W.",
    title = "{Convergence of Bogolyubov's method of renormalization in momentum space}",
    doi = "10.1007/BF01645676",
    journal = "Commun. Math. Phys.",
    volume = "15",
    pages = "208--234",
    year = "1969"
    }

@article{BP,
    author = "Bogoliubov, N. N. and Parasiuk, O. S.",
    title = "{On the Multiplication of the causal function in the quantum theory of fields}",
    doi = "10.1007/BF02392399",
    journal = "Acta Math.",
    volume = "97",
    pages = "227--266",
    year = "1957"
}

@article{APZ,
    author = "Anikin, S. A. and Polivanov, M. K. and Zavyalov, O. I.",
    title = "{SIMPLE PROOF OF THE BOGOLYUBOV-PARASIUK THEOREM}",
    reportNumber = "JINR-E2-7433",
    doi = "10.1007/BF01037256",
    journal = "Theor. Math. Phys.",
    volume = "17",
    pages = "1082",
    year = "1973"
}

@book{Elvang,
    author = "Elvang, Henriette and Huang, Yu-tin",
    title = "{Scattering Amplitudes in Gauge Theory and Gravity}",
    publisher = "Cambridge University Press",
    month = "4",
    year = "2015",
    address   = "Cambridge"
}

@article{1001,
    author = "Weinzierl, Stefan",
    title = "{Tales of 1001 Gluons}",
    eprint = "1610.05318",
    archivePrefix = "arXiv",
    primaryClass = "hep-th",
    reportNumber = "MITP-16-109",
    doi = "10.1016/j.physrep.2017.01.004",
    journal = "Phys. Rept.",
    volume = "676",
    pages = "1--101",
    year = "2017"
}

@manual{Schwinn,
  title        = "Modern Methods of Quantum Chromodynamics",
  author       = "Schwinn, Christian",
  organization = "Albert-Ludwigs-Universitat Freiburg, Physikalisches Institut
D-79104",
  address      = "Freiburg, Germany",
  year         = 2015
}

@article{Borlakov,
    author = "Borlakov, Arthur and Kazakov, Dmitry",
    title = "{Loop corrections to the four-fermion interaction}",
    eprint = "2504.03269",
    archivePrefix = "arXiv",
    primaryClass = "hep-th",
    doi = "10.1140/epjc/s10052-025-14733-6",
    journal = "Eur. Phys. J. C",
    volume = "85",
    number = "9",
    pages = "1000",
    year = "2025"
}

@article{Borlakov2,
    author = "Borlakov, Arthur and Kazakov, Dmitry",
    title = "{High-energy behaviour of Fermi theory}",
    eprint = "2511.22241",
    archivePrefix = "arXiv",
    primaryClass = "hep-th",
    doi = "10.1140/epjc/s10052-026-15317-8",
    journal = "Eur. Phys. J. C",
    volume = "86",
    number = "2",
    pages = "159",
    year = "2026"
}

@book{peskin,
  title={An Introduction To Quantum Field Theory},
  author={Peskin, M.E. and Schroeder, D.V.},
  year={2018},
  publisher={Ch 7.3. CRC Press}
}

@book{Schwartz,
    author = "Schwartz, Matthew D.",
    title = "{Quantum Field Theory and the Standard Model}",
    isbn = "978-1-107-03473-0, 978-1-107-03473-0",
    publisher = "Cambridge University Press",
    month = "3",
    year = "2014"
}

@book{GG,
  title={Gauge Theory of Weak Interactions},
  author={Greiner, W. and M{\"u}ller, B.},
  isbn={9783540879244},
  year={2009},
  publisher={Springer-Verlag Berlin Heidelberg}
}

\end{document}